\pdfoutput=1
\documentclass[a4paper,fleqn]{cas-dc}

\input{glyphtounicode}
\usepackage[numbers]{natbib}
\usepackage{amsmath,amssymb}
\usepackage{algorithm}
\usepackage{algpseudocode}
\usepackage{booktabs}
\usepackage{graphicx}
\usepackage{hyperref}
\usepackage{caption}

\begin{document}
\let\WriteBookmarks\relax
\def\floatpagepagefraction{1}
\def\textpagefraction{.001}

\shorttitle{BLANC: Blank Landscape Analysis through NPMI Conditioning}
\shortauthors{S. Miyazawa and K. Fujii}

\title[mode=title]{BLANC: Discovering Patent White Space via Changes in Normalized Pointwise Mutual Information Between Multi-View Clusters}

  \author[1]{Shuichi Miyazawa}
  \cormark[1]
  \ead{shuichi.miyazawa@agc.com}
  \author[2]{Kensuke Fujii}
  \ead{kensuke.fujii@agc.com}
  \affiliation[1]{organization={Digital \& Innovation Management Division, AGC Inc.},
                  city={Tokyo}, country={Japan}}
  \affiliation[2]{organization={Intellectual Property Division, AGC Inc.},
                  city={Tokyo}, country={Japan}}
  \cortext[1]{Corresponding author.}

\begin{abstract}
Identifying \emph{white space}, the unexplored but potentially valuable regions of a patent landscape, is essential for strategic R\&D planning, yet existing methods either rely on manual patent mapping or apply single-view clustering without quantitative gap detection.
We propose BLANC (Blank Landscape Analysis through NPMI Conditioning), a three-phase pipeline that combines (1)~multi-view neural topic modeling along three semantic dimensions (application/use, novelty, inventive step); (2)~Normalized Pointwise Mutual Information (NPMI) to quantify cross-dimensional cluster association; and (3)~conditional detection that flags combinations whose NPMI drops when the corpus is filtered by a user-specified keyword.
The drop is captured by a new metric, $\Delta\text{NPMI}$, which identifies combinations ``established globally, unexplored locally.''
Because white space has no ground truth in principle, we evaluate BLANC on two public USPTO corpora---machine learning/AI (5{,}417 patents, CPC G06N) and glass compositions (1{,}982 patents, CPC C03C)---by artificially depleting known technology combinations and testing whether BLANC recovers them.
When three-quarters of a target pair's documents are removed, BLANC recovers 34.1\% (ML/AI) and 27.3\% (glass) of the depleted combinations, whereas size-matched removals not aimed at them (random documents, or those of a \emph{different} established combination) essentially never do: the target is never recovered in 191 decoy trials. Collapsing the three semantic views into one recovers nothing, while prior co-occurrence measures also flag the target under random removal, offering no specificity.
Furthermore, a proprietary-data case (302 float glass / glass-ceramics patents) shows that, with the keyword ``fluorine,'' BLANC reveals a fluorine surface treatment $\times$ warpage suppression candidate ($\Delta\text{NPMI}$ up to 0.48) that experts had independently identified.
The method enables efficient, reproducible preparation for patent brainstorming in R\&D and IP departments.
\end{abstract}


\maketitle

\section{Introduction}\label{sec:intro}

In industrial R\&D and intellectual property (IP) departments, brainstorming sessions for new patent application themes are a critical activity.
However, the preparatory process of reading large volumes of related patents suffers from several inefficiencies:
\begin{enumerate}
\item reading patents without a clear objective is labor-intensive and difficult to motivate;
\item brainstorming quality depends heavily on individual domain expertise, leading to inconsistent outcomes across participants;
\item traditional patent maps\footnote{Manually constructed visual summaries of a patent landscape.} are time-consuming to create, require group effort, and often yield few actionable insights, since a blank region may be technologically meaningless; and
\item standard AI-based classification approaches do not incorporate an organization's own technical knowledge, yielding only generic perspectives.
\end{enumerate}

This paper addresses the problem of automatically finding \emph{``what should exist but doesn't''} in patent landscapes.
We propose BLANC (Blank Landscape Analysis through NPMI Conditioning), a system that discovers technology theme combinations that are well-represented in the broader patent population but absent or underrepresented when restricted to the user's keyword of interest.
These weakened associations represent candidate themes for new patent applications: the \emph{white space}.

Prior work approaches patent white space along three lines.
Early morphological and map-based methods flag empty cells in keyword or classification matrices~\cite{Yoon2005,Son2012,Yoon2018gtm};
cluster- and function-based methods infer gaps from inter-cluster connectivity or Subject--Action--Object (SAO) structures (subject--verb--object triples extracted from patent text)~\cite{Kim2014,Yoon2015function,Choi2014};
and recent neural approaches apply transformer-based topic models such as BERTopic~\cite{Grootendorst2022} to technology-opportunity discovery~\cite{Jeon2023,Yang2025}.
A parallel line quantifies how strongly technology pairs co-occur, using Pointwise Mutual Information (PMI) and its normalized form NPMI~\cite{Church1990,Bouma2009} or IPC co-classification statistics~\cite{Jaffe1986,Breschi2003}.

Yet no method combines the three ingredients we argue are jointly necessary: multi-view clustering, a normalized cross-dimensional co-occurrence measure, and keyword-conditional gap detection. The closest precursors each address only part of this: Kim et al.~\cite{Kim2014} use traditional clustering without a normalized association measure, while Jeon et al.~\cite{Jeon2023} use single-view clustering without statistical gap detection (Table~\ref{tab:gap}).
BLANC closes this gap.

The approach consists of three phases:

\textbf{Phase~1 (Multi-view clustering):}
Patent documents are independently clustered along three semantic dimensions: application/use, novelty, and inventive step. The BERTopic neural topic modeling pipeline performs the clustering, embedding documents with Sentence-BERT, reducing their dimensionality with UMAP, and grouping them with HDBSCAN.\footnote{UMAP: Uniform Manifold Approximation and Projection. HDBSCAN: Hierarchical Density-Based Spatial Clustering of Applications with Noise.}

\begin{figure*}[t]
\centering
\includegraphics[width=\textwidth]{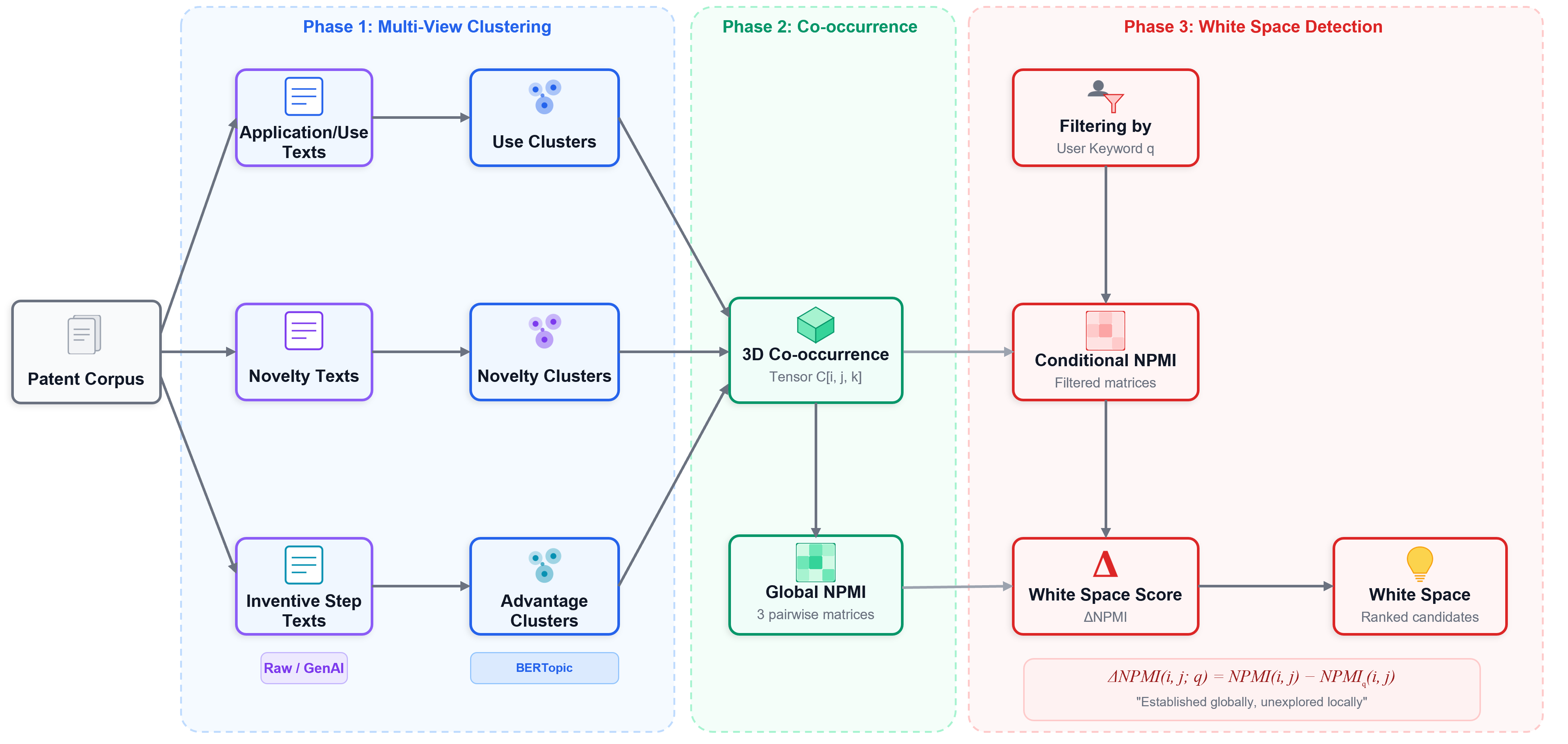}
\caption{System architecture and data flow of the BLANC pipeline. Phase~1 independently clusters patents along three semantic dimensions using BERTopic. Phase~2 constructs a 3D co-occurrence tensor and computes pairwise NPMI matrices. Phase~3 filters by a user keyword, recomputes conditional NPMI, and ranks cluster pairs by $\Delta\text{NPMI}$.}\label{fig:pipeline}
\end{figure*}

\textbf{Phase~2 (Co-occurrence quantification):}
Statistical association between clusters from different dimensions is computed via Normalized Pointwise Mutual Information (NPMI), producing pairwise NPMI matrices that capture strongly related theme combinations.

\textbf{Phase~3 (White space detection):}
When the user specifies a keyword, the corpus is filtered and conditional NPMI is recomputed.
Cluster pairs whose NPMI drops markedly are flagged as white space using a new metric, the NPMI drop ($\Delta\text{NPMI}$).

The contributions of this paper are threefold:
\begin{enumerate}
\item \textbf{Methodological:} We propose the BLANC pipeline, the first to integrate multi-view BERTopic clustering with NPMI co-occurrence analysis for patent white space detection.
\item \textbf{Conceptual:} We introduce the NPMI drop ($\Delta\text{NPMI}$), a metric that operationalizes white space as ``established globally, unexplored locally'' by computing the difference in NPMI between the full corpus and a keyword-filtered subcorpus.
\item \textbf{Practical:} We evaluate BLANC on two public USPTO corpora spanning distinct technology domains (ML/AI and glass compositions) under a circularity-robust, removal-driven protocol. BLANC recovers injected white space, whereas size-matched control removals (random documents, or the documents of a different established combination) essentially do not. Under the identical protocol, BLANC surpasses external prior methods in specificity and coverage: unnormalized inter-cluster connectivity~\cite{Kim2014} shows spurious recovery under random removal, and a CPC co-classification substitute~\cite{Breschi2003} exposes substantially fewer testable combinations. A proprietary case further illustrates recovery of an expert-identified candidate.
\end{enumerate}

\section{Related Work}\label{sec:related}

\begin{table*}[t]
\caption{Comparison with closest prior work. \checkmark\ = present; --- = absent.}\label{tab:gap}
\centering
\begin{tabular}{lcccc}
\toprule
Study & Clustering & Co-occ.\ measure & Multi-view & Cond.\ detection \\
\midrule
Kim et al.\ (2014)    & Traditional & Connectivity\textsuperscript{$\dagger$} & ---        & --- \\
Choi \& Jun (2014)    & Bayesian    & ---            & ---        & --- \\
Jeon et al.\ (2023)   & BERTopic    & ---            & ---        & --- \\
Yang \& Chen (2025)   & BERTopic    & ---            & ---        & --- \\
\textbf{This work}    & \textbf{BERTopic} & \textbf{NPMI} & \checkmark & \checkmark \\
\bottomrule
\multicolumn{5}{l}{\footnotesize $^{\dagger}$Unnormalized inter-cluster connectivity, not a normalized co-occurrence association measure.}
\end{tabular}
\end{table*}

The proposed system integrates techniques from three research streams.
This section reviews each in detail; the gap relative to the closest prior art was established in Section~\ref{sec:intro}.

\subsection{Neural topic modeling and patent-specific embeddings}\label{sec:rw-nlp}

BERTopic~\cite{Grootendorst2022} is a modular framework that generates document embeddings with pre-trained transformers, reduces dimensionality with UMAP~\cite{McInnes2018}, clusters with HDBSCAN~\cite{McInnes2017}, and produces topic representations via class-based Term Frequency--Inverse Document Frequency (c-TF-IDF).
It consistently outperforms Latent Dirichlet Allocation (LDA)~\cite{Blei2003} and Non-negative Matrix Factorization (NMF)~\cite{LeeSeung1999} on coherence and diversity benchmarks.
For patent text, domain-adapted embeddings improve clustering quality: Sentence-BERT~\cite{Reimers2019} enables large-scale semantic clustering, while PatentBERT~\cite{Lee2020patentbert} and PatentSBERTa~\cite{Bekamiri2024} further specialize embeddings to the patent domain.
Recent surveys document the transition to patent analysis based on large language models (LLMs) and note that such applications remain underexplored~\cite{Krestel2021,Jiang2025survey}.

\subsection{PMI, NPMI, and co-occurrence measures in patent analysis}\label{sec:rw-pmi}

PMI~\cite{Church1990} quantifies how much the co-occurrence of two events exceeds independence, and its normalized form NPMI~\cite{Bouma2009} scales values to $[-1, +1]$, with $-1$ indicating complete absence of co-occurrence, $0$ independence, and $+1$ perfect co-occurrence.
This bounded normalization is critical for comparing association strengths across cluster pairs of different frequencies.
In technology mapping, Jaffe's~\cite{Jaffe1986} seminal use of patent classification vectors established the economic rationale for co-occurrence-based proximity.
Breschi et al.~\cite{Breschi2003} and Curran and Leker~\cite{Curran2011} formalized International Patent Classification (IPC) co-classification as a test of whether technology pairs co-occur more than by chance.
This is the same logic that NPMI applies to cluster pairs.

\subsection{Patent white space analysis and technology gap detection}\label{sec:rw-ws}

White space analysis has evolved through three generations.
Morphological approaches~\cite{Yoon2005,Lee2009} organize patent keywords into 2D maps (morphology matrices or principal component analysis projections) and flag unoccupied cells or sparse regions as unexplored combinations.
Son et al.~\cite{Son2012} and Yoon and Magee~\cite{Yoon2018gtm} advanced vacancy detection with Generative Topographic Mapping, shifting from detection to probabilistic prediction.
Cluster-based and function-based methods followed.
Kim et al.~\cite{Kim2014} measured inter-cluster connectivity on text and citation networks; this is the closest precursor to computing co-occurrence between cluster pairs, although it uses traditional clustering without normalized association measures.
Yoon et al.~\cite{Yoon2015function} used SAO structures to identify unexploited technology--product pairings, and Choi and Jun~\cite{Choi2014} applied Bayesian clustering for the same goal.
Lee and Lee~\cite{Lee2019recombinant} grounded white space analysis in recombinant search theory, which holds that innovation arises from novel combinations of existing components.
Transformer-based work is more recent.
Jeon et al.~\cite{Jeon2023} applied BERTopic with PatentSBERTa to USPTO data for technology opportunity discovery; this is the most directly analogous prior work, although it uses single-view clustering without statistical co-occurrence measures.
Yang and Chen~\cite{Yang2025} integrated BERTopic with logistic growth models for lifecycle analysis.
Across these generations, however, the same documents have not been clustered along several independent semantic dimensions; a plausible reason is that obtaining perspective-separated patent texts at scale was, until recently, impractical. The structured fields exposed by modern patent corpora (e.g., abstract, claims, and summary) now make this feasible; where such fields are absent, generative-AI extraction from the full text serves the same purpose.

\section{Proposed Method: BLANC}\label{sec:method}

Fig.~\ref{fig:pipeline} shows the BLANC architecture.
The pipeline processes a patent corpus through three phases: multi-view clustering, co-occurrence quantification, and conditional white space detection.
Each patent contributes three dimension-specific texts (application/use, novelty, and inventive step), taken from native structured fields (e.g., abstract, claims, and summary) or, where these are absent, extracted by a generative AI model; field-specific preprocessing of these texts and the document-term matrix (DTM) used for keyword filtering in Phase~3 are detailed in Appendix~\ref{app:public}.

\subsection{Multi-view clustering via BERTopic}\label{sec:clustering}

The key idea is to treat different text fields as independent semantic views, analogous to multi-view learning, where the same dataset is analyzed from multiple representations independently.
We define $K=3$ dimensions: (1)~application/use text, capturing \emph{where} the invention is used; (2)~novelty text, capturing \emph{what} is new; and (3)~inventive step text, capturing \emph{what benefit} it provides.
These three views are not arbitrary, but mirror the canonical structure of a patent specification. Its statement of industrial applicability describes \emph{where} the invention is used, its claims describe \emph{what} is new, and its account of the technical effect or problem solved describes \emph{why} it is useful. Each view therefore corresponds to an established documentary distinction rather than an ad-hoc split.
Other decompositions (e.g., technical field, problem addressed) are possible; we adopt $K=3$ for this correspondence and leave a systematic study of the number and choice of views to future work (\S\ref{sec:disc-limit}).

For each dimension $k \in \{1, 2, 3\}$, the BERTopic pipeline is executed independently:
\begin{enumerate}
\item \textbf{Embedding:} Dimension-specific text is encoded into dense vectors using a pre-trained Sentence-BERT model~\cite{Reimers2019}.
\item \textbf{Dimensionality reduction:} UMAP~\cite{McInnes2018} projects embeddings into a lower-dimensional space.
\item \textbf{Clustering:} HDBSCAN~\cite{McInnes2017} identifies dense clusters; documents not fitting any cluster are classified as noise (cluster ID $= -1$).
\item \textbf{Topic representation:} Each cluster is assigned representative keywords via a composite scoring method (Appendix~\ref{app:public}).
\end{enumerate}

Each document $d$ is thus assigned a cluster triple $d \to (c_1, c_2, c_3)$, where $c_k$ is the cluster assignment for dimension $k$, or $-1$ if the document is noise in that dimension.

\begin{algorithm*}[t]
\caption{BLANC: Patent White Space Detection Pipeline}\label{alg:ws}
\begin{algorithmic}[1]
\Require Patent corpus $D$, semantic dimensions $k \in \{1,\ldots,K\}$, association threshold $\theta_{\min}$ (default 0.3)
\Ensure Ranked white space candidates
\For{each dimension $k$}
  \State Extract dimension-specific texts (raw fields or generative AI)
  \State Run BERTopic: Sentence-BERT $\to$ UMAP $\to$ HDBSCAN $\to$ Keywords
  \State Assign cluster labels $c_k(d)$ for all $d \in D$
\EndFor
\State Construct 3D co-occurrence tensor $C[i,j,k]$ \Comment{Eq.~(\ref{eq:tensor})}
\State Marginalize to three 2D matrices; compute global NPMI \Comment{Eqs.~(\ref{eq:marginal})--(\ref{eq:npmi})}
\State \textbf{User} inspects the global NPMI landscape and specifies a keyword $q$ of interest \Comment{interactive}
\State Filter corpus: $D_q \gets \{d \mid \mathrm{DTM}[d,q] > 0\}$
\State Compute conditional tensor and $\mathrm{NPMI}_q$ over $D_q$
\For{each cluster pair $(i,j)$ across dimension combinations}
  \State $\Delta\text{NPMI}(i,j;q) \gets \mathrm{NPMI}(i,j) - \mathrm{NPMI}_q(i,j)$ \Comment{Eq.~(\ref{eq:ws})}
  \If{$\Delta\text{NPMI} > 0$ \textbf{and} $\mathrm{NPMI}(i,j) \geq \theta_{\min}$ \textbf{and} pair present in $D_q$ (co-occurrence $\geq 1$)}
    \State Add $(i,j)$ to candidate list with $\Delta\text{NPMI}(i,j;q)$
  \EndIf
\EndFor
\State \Return the top-20 candidates per dimension combination, ranked by $\Delta\text{NPMI}$
\Statex \hrulefill
\Statex \textbf{Micro-level drill-down} (optional, per candidate):
\State Extract document subset matching candidate cluster pair $(i,j)$
\State Optionally intersect with $D_q$ for keyword-focused analysis
\State Build phrase co-occurrence matrix; compute pairwise NPMI
\State Cluster phrases via Ward linkage on $1 - \mathrm{NPMI}$ distance
\State Rank phrases by $\mathrm{NPMI}(w, q)$ if keyword filter is active
\end{algorithmic}
\end{algorithm*}

\subsection{Co-occurrence quantification: 3D tensor and NPMI}\label{sec:tensor}

With $K=3$ dimensions having $N_1$, $N_2$, $N_3$ clusters respectively, we construct a 3D co-occurrence tensor:
\begin{equation}\label{eq:tensor}
C[i, j, k] = \bigl| D_i^{(1)} \cap D_j^{(2)} \cap D_k^{(3)} \bigr|
\end{equation}
counting documents simultaneously assigned to cluster $i$ in dimension~1, cluster $j$ in dimension~2, and cluster $k$ in dimension~3.
Documents that HDBSCAN leaves unassigned (noise, label $-1$) in \emph{any} dimension are excluded from the tensor. A pairwise count obtained by marginalization can therefore be smaller than the number of documents sharing that cluster pair in the two dimensions concerned, since the latter does not constrain the third dimension.
For pairwise analysis, the tensor is marginalized over each dimension to produce three 2D matrices (e.g., $C^{(1 \cdot 2)}[i,j] = \sum_k C[i,j,k]$).

For each 2D co-occurrence matrix $C$ of size $M \times N$, NPMI quantifies the association strength between cluster pairs.
Let $T = \sum_{i,j} C_{ij}$ be the total co-occurrences.
Then:
\begin{align}
P(x_i) &= \sum_j C_{ij} / T, \quad P(y_j) = \sum_i C_{ij} / T \label{eq:marginal}\\
P(x_i, y_j) &= C_{ij} / T \label{eq:joint}\\
\mathrm{PMI}(x_i, y_j) &= \log_2 \frac{P(x_i, y_j)}{P(x_i) \cdot P(y_j)} \label{eq:pmi}\\
\mathrm{NPMI}(x_i, y_j) &= \frac{\mathrm{PMI}(x_i, y_j)}{-\log_2 P(x_i, y_j)} \label{eq:npmi}
\end{align}
where a small constant is added to each probability for numerical stability.

\subsection{Conditional white space detection}\label{sec:whitespace}

White space detection flags cluster pairs whose NPMI is high in the full corpus but drops when the corpus is restricted to documents containing a user-specified keyword.

Given a user keyword $q$ and the DTM, the filtered corpus is $D_q = \{d \mid \mathrm{DTM}[d, q] > 0\}$.
The conditional co-occurrence tensor and conditional NPMI values $\mathrm{NPMI}_q(i,j)$ are computed over $D_q$ using the same formulas as Eqs.~(\ref{eq:marginal})--(\ref{eq:npmi}).
We quantify white space for a cluster pair $(i,j)$ by the \emph{NPMI drop}, denoted $\Delta\text{NPMI}$:
\begin{equation}\label{eq:ws}
\Delta\text{NPMI}(i, j; q) = \mathrm{NPMI}(i, j) - \mathrm{NPMI}_q(i, j)
\end{equation}
A positive value indicates that the combination is underrepresented among patents containing keyword $q$; that is, it is established globally but unexplored locally.

The metric is retained only if: (a)~the filtered corpus contains at least one document assigned to both clusters \emph{jointly}, i.e.\ conditional co-occurrence $\geq 1$; (b)~$\Delta\text{NPMI} > 0$; and (c)~the original $\mathrm{NPMI}(i,j) \geq \theta_{\min}$ (default: 0.3).
\begin{figure*}[t]
\centering
\includegraphics[width=\textwidth]{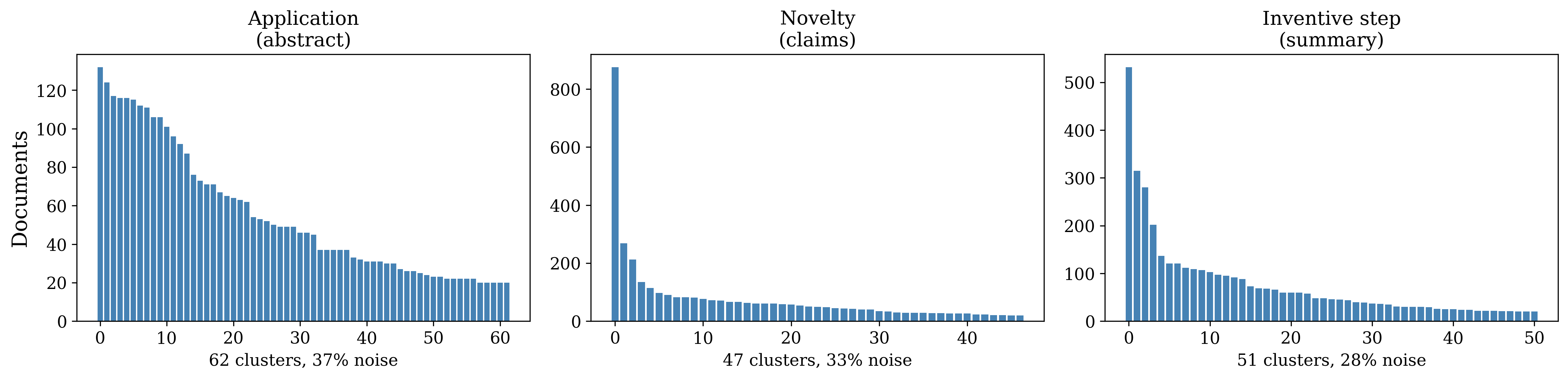}
\caption{Cluster size distribution for the G06N corpus after field-specific preprocessing. Each dimension resolves into many clusters (62/47/51); the application dimension is finely graded, while novelty and inventive step each retain one larger leading cluster but still spread across dozens of themes rather than collapsing into a single dominant cluster as they do without preprocessing, enabling meaningful cross-dimensional NPMI analysis. Note: Y-axis scales differ across panels to accommodate the different cluster size ranges.}\label{fig:cluster-sizes}
\end{figure*}

Condition~(c) is essential.
For a pair whose global NPMI is already low, $\Delta\text{NPMI}$ can turn positive under \emph{any} unrelated keyword, since the filtered corpus retains only loosely related documents, which mechanically weakens an already weak association---a trivial gap, not white space. Restricting to pairs above $\theta_{\min}$ isolates only the \emph{surprising} cases, i.e., pairs that are strongly coupled across the full corpus yet decoupled once the corpus is narrowed to the keyword.
The top-20 pairs from each dimension combination are presented as candidates.

Algorithm~\ref{alg:ws} summarizes the full pipeline.

\subsection{Micro-level phrase analysis (optional)}\label{sec:micro}

In addition to the macro-level cluster-pair analysis, BLANC offers an optional drill-down for inspecting individual candidates.
Given a candidate pair $(i, j)$, the subset of documents satisfying both cluster constraints is extracted, and a phrase co-occurrence matrix is built from the most frequent phrases, on which pairwise NPMI is computed.
Hierarchical agglomerative clustering on the $1 - \mathrm{NPMI}$ distance, using Ward linkage~\cite{Ward1963} (the criterion that merges clusters to minimize the increase in within-cluster variance), reveals \emph{phrase theme groups}; the heatmap is reordered by group assignment to visualize within-group coherence and between-group separation.
When a keyword filter $q$ is active, phrases are additionally ranked by $\mathrm{NPMI}(w, q)$ computed with the same formula as Eq.~(\ref{eq:npmi}) but over phrase--keyword co-occurrence. This ranking shows which phrases within the underexplored combination are most associated with the keyword of interest.

\textbf{Computational complexity.}
Post-embedding operations are inexpensive: tensor construction is $O(N_1 N_2 N_3 |D|)$, NPMI is $O(MN)$ per 2D matrix, and keyword filtering is $O(|D|)$.
For typical corpora of thousands of documents and tens of clusters per dimension, all post-embedding steps complete in seconds; the BERTopic embedding step is the dominant cost and scales linearly with corpus size.

\section{Experiments}\label{sec:experiments}

We evaluate BLANC on two public USPTO corpora and one proprietary corpus, covering clustering (\S\ref{sec:res-clustering}), global NPMI (\S\ref{sec:res-npmi}), white space case studies (\S\ref{sec:res-ws}), industrial validation on proprietary data (\S\ref{sec:res-proprietary}), and baselines (\S\ref{sec:eval-baseline}); a theme-keyword sensitivity diagnostic is reported in Appendix~\ref{sec:eval-synthetic}.
All software, hyperparameters, preprocessing, and reproducibility settings are detailed in Appendix~\ref{app:public}.

\subsection{Dataset}\label{sec:data}

We use the Harvard USPTO Patent Dataset (HUPD)~\cite{Suzgun2024hupd}, selecting two technologically distinct Cooperative Patent Classification (CPC) subclasses to evaluate cross-domain generalizability:

The \textbf{G06N corpus} contains 5{,}417 applications in CPC subclass G06N (computing arrangements based on specific computational models), covering machine learning, neural networks, and genetic algorithms, filed 2013--2016.
The \textbf{C03C corpus} contains 1{,}982 applications in CPC subclass C03C (chemical composition of glasses, glazes, or vitreous enamels), covering optical glass, glass fiber, coatings, and chemical strengthening, filed 2012--2017.
For both corpora, abstracts, claims, and summaries are mapped to the application, novelty, and inventive step dimensions respectively (Table~\ref{tab:data}).

On the G06N corpus, without the field-specific preprocessing detailed in Appendix~\ref{app:public}, claims collapse to 2~clusters and summaries to 8~clusters (one dominant cluster holding 80\% of documents); after preprocessing they produce 47 and 51~clusters (\S\ref{sec:eval-sensitivity}).
DTM construction applies a patent-domain stop-word list, in addition to the standard English stop words, to remove non-discriminative terms.
The C03C corpus yields a larger DTM vocabulary (80{,}244 vs.\ 65{,}606) despite fewer documents, reflecting rich chemical nomenclature such as ``SiO$_2$--Al$_2$O$_3$--Li$_2$O''.

\begin{table}[t]
\caption{Dataset statistics for the two HUPD corpora.}\label{tab:data}
\centering
\small
\begin{tabular}{lcc}
\toprule
 & G06N & C03C \\
 & (ML/AI) & (Glass) \\
\midrule
Documents     & 5{,}417 & 1{,}982 \\
Filing years        & 2013--16 & 2012--17 \\
Avg.\ words (abstract)       & 114 & 100 \\
Avg.\ words (claims)     & 1{,}216 & 764 \\
Avg.\ words (summary)  & 684 & 729 \\
Empty rate (summary)      & 1.6\% & 14.9\% \\
DTM vocabulary      & 65{,}606 & 80{,}244 \\
\bottomrule
\end{tabular}
\end{table}

\subsection{Clustering results}\label{sec:res-clustering}

\begin{figure*}[t]
\centering
\includegraphics[width=\textwidth]{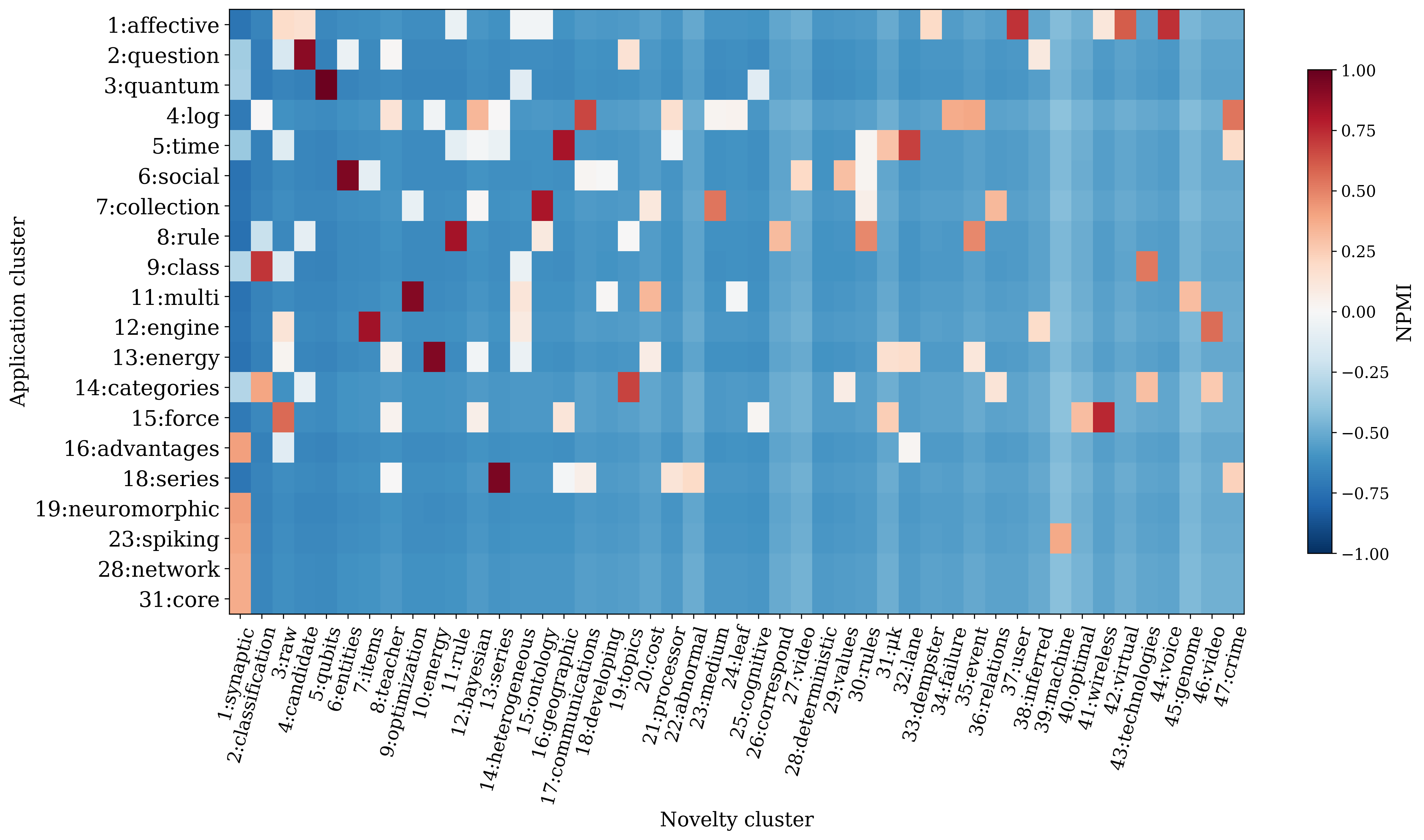}
\caption{NPMI heatmap for application$\times$novelty cluster pairs (G06N corpus, top-20 application clusters by co-occurrence; vertical-axis numbers are the original cluster IDs and are therefore non-consecutive, since only these top-20 clusters are shown). Dark-red cells mark cluster pairs whose application and novelty themes co-occur far more than chance (coherent technology combinations), whereas most cells are weak (blue), indicating that the majority of cross-dimensional combinations are only loosely associated.}\label{fig:npmi-heatmap}
\end{figure*}
For G06N, BERTopic identified 62, 47, and 51~clusters for the application, novelty, and inventive step dimensions (noise rates 36.8\%, 33.2\%, 28.3\%); the comparatively high application-dimension rate reflects HDBSCAN's intentional exclusion of documents too short or too generic to form coherent clusters.
The application dimension spans diverse themes including quantum computing, question answering, social networking, autonomous driving, anomaly detection, spiking neural networks, sentiment analysis, and agricultural AI.
Fig.~\ref{fig:cluster-sizes} shows the resulting size distributions, which are finely graded in the application dimension, with a larger leading cluster in novelty and inventive step but no collapse into a single dominant cluster.
For C03C, BERTopic produced 63, 63, and 48~clusters (noise rates 21.3\%, 21.3\%, 17.3\%), confirming that the pipeline scales to a substantially different technology domain.

\subsection{Co-occurrence and global NPMI}\label{sec:res-npmi}

The application$\times$novelty NPMI matrix ($62 \times 47$) reveals the associative structure of the ML/AI landscape; Fig.~\ref{fig:npmi-heatmap} visualizes the top-20 application rows and Table~\ref{tab:npmi-top} lists the top-10 pairs by NPMI.

The strongest association (NPMI of 0.978) is between quantum computing clusters in both dimensions, reflecting the tight coupling between quantum processor applications and quantum-specific technical claims.
Other highly associated pairs include decision trees (NPMI of 0.958), finite automata (0.948), time series forecasting (0.939), and social networking (0.930).
These high-NPMI pairs represent well-established technology combinations in the ML/AI domain.

\begin{table}[t]
\caption{Top-10 app$\times$novelty pairs by global NPMI. Co.: co-occurrence count.}\label{tab:npmi-top}
\centering
\footnotesize
\begin{tabular}{llcr}
\toprule
Application theme & Novelty theme & NPMI & Co. \\
\midrule
Quantum comp.   & Quantum proc. & 0.978 & 106 \\
Decision tree    & Decision tree & 0.958 &  33 \\
Finite automaton & DFA/NFA      & 0.948 &  30 \\
Time series      & Forecast model & 0.939 &  44 \\
Social network   & Social graph  & 0.930 &  61 \\
Energy cons.     & Energy anomaly & 0.926 &  60 \\
User expertise   & Expertise inf. & 0.919 &   8 \\
Multi-obj opt.   & Pareto sol.    & 0.916 &  55 \\
Question ans.    & Cand.\ answers & 0.906 &  83 \\
Anomaly det.     & Anomaly score  & 0.878 &  26 \\
\bottomrule
\end{tabular}
\end{table}

\subsection{White space detection: case studies}\label{sec:res-ws}

To demonstrate the white space detection capability, we selected the keyword ``neural'' as the user query, targeting neural network (NN) technologies within the broader ML/AI corpus.
Filtering the corpus by this keyword reduced the document count from 5{,}417 to 1{,}014 (18.7\% of the original).

\begin{figure*}[t]
\centering
\includegraphics[width=0.85\textwidth]{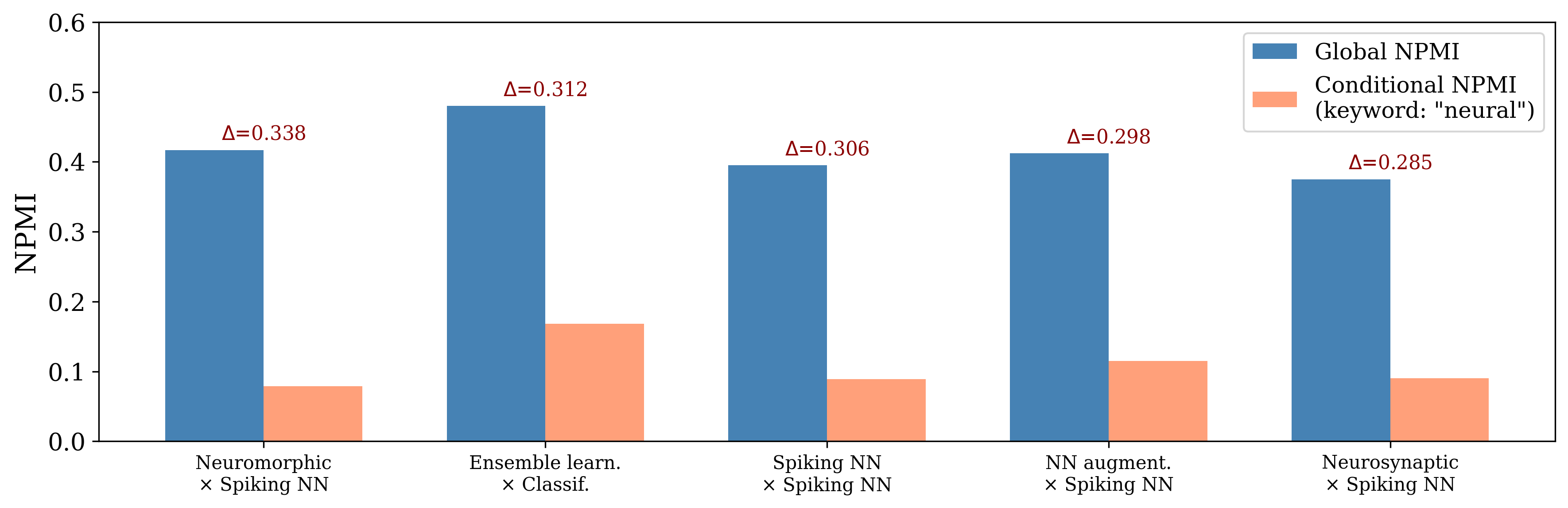}
\caption{Top-5 white space candidates for keyword ``neural'' (G06N, application$\times$novelty; same candidates as Table~\ref{tab:ws}). Blue bars show global NPMI; salmon bars show conditional NPMI after keyword filtering. The gap ($\Delta\text{NPMI}$, annotated) represents the degree of underrepresentation among ``neural''-keyword patents.}\label{fig:delta-npmi-bar}
\end{figure*}

Fig.~\ref{fig:delta-npmi-bar} and Table~\ref{tab:ws} present the top white space candidates ranked by $\Delta\text{NPMI}$ for the application$\times$novelty dimension pair.

The top candidate pairs neuromorphic hardware (application cluster~19: ``neuromorphic synapse'') with spiking NN claims (novelty cluster~1: ``synaptic / output neuron''); its global NPMI of 0.417 falls to a conditional NPMI of 0.079, an NPMI drop of 0.338. This combination is well established overall but underrepresented among ``neural''-keyword patents, since neuromorphic patents emphasize circuit-level rather than neural-network framing.

Four of five candidates share the spiking-NN novelty cluster, pairing it with distinct application clusters: neuromorphic hardware, spiking networks, NN augmentation, and neurosynaptic cores.
The fifth (ensemble learning $\times$ classification) rests on a single co-occurring document ($n_q=1$) and is reported only for completeness.
This concentration identifies spiking/neuromorphic computing as a coherent area systematically underrepresented in ``neural''-keyword patents. The gap is terminological, since neuromorphic patents tend to use ``synaptic'', ``spike'', or ``membrane''. This shows that BLANC can surface non-obvious vocabulary-divergence gaps that support brainstorming.

\begin{table}[t]
\caption{Top-5 white space candidates for keyword ``neural'' (app$\times$novelty). $\mathrm{NPMI}$: global association; $\mathrm{NPMI}_q$: conditional association on the keyword-filtered subcorpus; $\Delta = \mathrm{NPMI} - \mathrm{NPMI}_q$; $n_q$: conditional co-occurrence count.}\label{tab:ws}
\centering
\footnotesize
\begin{tabular}{llcccc}
\toprule
Application & Novelty & $\mathrm{NPMI}$ & $\mathrm{NPMI}_q$ & $\Delta$ & $n_q$ \\
\midrule
Neuromorphic     & Spiking NN   & 0.417 & 0.079 & 0.338 & 19 \\
Ensemble learn.  & Classif.     & 0.480 & 0.168 & 0.312 & 1 \\
Spiking NN       & Spiking NN   & 0.395 & 0.089 & 0.306 & 37 \\
NN augment.      & Spiking NN   & 0.412 & 0.115 & 0.298 & 68 \\
Neurosynaptic    & Spiking NN   & 0.375 & 0.090 & 0.285 & 28 \\
\bottomrule
\end{tabular}
\end{table}

\textbf{C03C case study.}
The keyword ``abbe'' targets optical glass via the Abbe number, a measure of optical dispersion, and filters C03C to 50~documents (2.5\%).
BLANC returns two high-global/near-zero-conditional candidates: optical glass $\times$ optical glass molding ($\Delta\text{NPMI}$ of 0.770) and optical glass $\times$ La/Gd rare-earth composition ($\Delta\text{NPMI}$ of 0.654).
Both indicate that ``abbe'' patents focus on optical-property specification rather than molding or rare-earth composition. This is a potentially actionable gap for optical glass manufacturers.

\subsection{Industrial validation on proprietary data}\label{sec:res-proprietary}

We further apply BLANC to a proprietary corpus of 302~patents in float glass / glass-ceramics, filed 2019--2023.
Unlike HUPD, each document carries pre-structured fields (application, solution, problem) mapped directly to the three dimensions; embeddings use a multilingual Sentence-BERT model, with UMAP (\texttt{n\_neighbors=10}, \texttt{n\_components=5}) and HDBSCAN (\texttt{min\_cluster\_size=8}, \texttt{min\_samples=3}) on application and inventive step dimensions.
The novelty dimension required a $k$-means ($k=10$) fallback because HDBSCAN produced only 2~clusters.
BLANC yielded 15, 10, and 11~clusters (noise rates 12.6\%, 0\%, 15.9\%) spanning themes such as coating, laminated glass, solar cells, float strengthening, strengthening/warpage suppression, lithium aluminosilicate (LAS) crystallization, and cooking appliances.
$\theta_{\min}$ was relaxed to 0.2 because the default 0.3 leaves too few candidates on this small corpus.

Prior to the analysis, IP experts had already identified ``fluorine surface treatment'' $\times$ ``warpage suppression'' as an underexplored filing opportunity.
With the keyword ``fluorine'' (14~matches, 4.6\%), BLANC returned the flat/float-strengthened $\times$ strengthening/warpage suppression pair as the top candidate in all three dimension combinations: $\Delta\text{NPMI}$ of 0.464 for application$\times$novelty and 0.480 for application$\times$inventive step.
BLANC thus identified the same combination the experts had flagged. Because that pair was known to exist beforehand, we present this as a \emph{confirmatory} single case demonstrating consistency with expert judgment, not a blinded test of independent discovery.
The conditional estimate rests on only 14~documents (a caveat revisited in \S\ref{sec:disc-limit}); the candidate is nonetheless consistent across all three dimension pairs.
The micro-level phrase analysis (\S\ref{sec:micro}) further confirms the gap.
Phrases around warpage control (chemical strengthening, ion exchange, stress profile) and around fluorine treatment (surface treatment, fluorination, concentration profile) form two weakly connected theme groups.

\subsection{Quantitative evaluation}\label{sec:eval}

\subsubsection{Sensitivity to preprocessing and hyperparameters}\label{sec:eval-sensitivity}

The 2$\to$47 and 8$\to$51 cluster jumps reported in Section~\ref{sec:data} for claims and summaries respectively quantify the impact of field-specific preprocessing.
Without boilerplate removal, the claims and summary dimensions collapse to a handful of clusters dominated by legal formulaic language and figure-reference boilerplate, and multi-view analysis becomes impossible.
A systematic sensitivity analysis of hyperparameters ($\theta_{\min}$, HDBSCAN \texttt{min\_cluster\_size}, the keyword-scoring weight $\alpha$ defined in Appendix~\ref{app:public}, the number of views $K$, and embedding model choice) is deferred to future work; the removal-driven recovery results in Section~\ref{sec:eval-removal} provide indirect evidence of robustness, as a consistent targeted-versus-random separation was obtained with a single parameter configuration across two technologically distinct corpora.

\subsubsection{White space injection by removal-driven recovery}\label{sec:eval-removal}

Because no ground-truth white space labels exist for patent corpora, we evaluate by injection, artificially depleting a known combination and testing whether BLANC recovers it.
To make \emph{recovery} the quantity under test, rather than keyword targeting, we condition on a keyword that is \emph{independent} of the target pair's specific theme, so the pair is not flagged before injection and the only mechanism that can surface it is the removal itself.
(The alternative of matching the keyword to the target pair is confounded by the keyword itself; we analyze it separately as a metric-sensitivity diagnostic in Appendix~\ref{sec:eval-synthetic}.)

The protocol proceeds in four steps.
\begin{enumerate}
\item \textbf{Select established combinations.} We consider each application$\times$novelty pair with global NPMI~$\geq\theta_{\min}$ and co-occurrence~$\geq 12$.
\item \textbf{Assign an independent keyword.} For each pair we select a keyword by a fixed, \emph{theme-blind} rule that touches the pair's documents only through a coverage constraint.
Candidates are the master-DTM terms of at least three characters that occur in at least 10\% of the corpus and in at least 60\% of the pair's documents.\footnote{Varying this coverage threshold to 50\% or 70\% leaves the qualitative result intact: targeted recovery at $\delta=0.75$ is 32.6\%/34.1\%/37.2\% on G06N ($n=43/44/43$) and 26.1\%/27.3\%/9.1\% on C03C ($n=23/22/22$) at 50\%/60\%/70\%, with all controls at 0\% throughout; the eligible set shifts with the threshold. The C03C rate is nevertheless clearly threshold-sensitive, which reinforces that these rates are indicative rather than precise. Keyword concentration is likewise stable: ``computer'' takes 39, 37 and 32 of the G06N targets at the three thresholds.}
The master DTM contains unigrams to trigrams over the two text fields, with English and patent-domain stop words removed, $\mathrm{df}\geq 5$ on G06N and $\geq 3$ on C03C, and $\mathrm{df}\leq 80\%$ (Appendix~\ref{app:public}).
Among these candidates we take the term with the highest \emph{corpus-wide} document frequency, breaking ties alphabetically.
``The pair's documents'' here means every document assigned to both clusters in the application and novelty dimensions, leaving the third dimension unconstrained (\S\ref{sec:tensor}).
The coverage constraint is what guarantees that removing documents from pair~$\cap\,D_q$ actually depletes the pair \emph{inside} the filtered corpus, while ranking by corpus-wide frequency makes the term as generic as the constraint allows.
The rule is not injective, and in practice it returns a handful of common words unrelated to any pair's specific theme (``computer'' for 37 of the 44 G06N pairs; ``surface'' or ``temperature'' for 15 of the 22 C03C pairs; Appendix~\ref{app:keywords}).
\item \textbf{Keep only initially undetected pairs.} Because such a broad term spans many clusters, the target pair is only a small fraction of $D_q$ and its marginals are not inflated, so the pair lies outside the top-20 \emph{before} any removal. We retain only these initially undetected pairs: on G06N, 46 high-NPMI pairs yield 45 with a qualifying keyword and 44 initially undetected; on C03C, 23 yield 23 and 22.
Across the retained pairs the target accounts for a median 0.56\% of $D_q$ (maximum 7.0\%) on G06N and 1.5\% (maximum 3.3\%) on C03C, confirming that the keyword does not preferentially select the pair.
\item \textbf{Inject and measure recovery.} We remove a fraction~$\delta$ of the documents in pair~$\cap\,D_q$ and record whether the pair newly enters the top-20. Each run is matched by three controls that remove the same number of documents under the same fixed-seed protocol, differing only in \emph{which} documents are removed:
(i)~\emph{corpus-random}, drawn uniformly from the entire corpus, and
(ii)~\emph{$D_q$-random}, drawn uniformly from $D_q$ excluding the pair.
The third control, (iii)~\emph{decoy}, removes documents of a \emph{different} established pair inside the same $D_q$---the one whose $|\text{pair}\cap D_q|$ is closest to the target's among those still undetected before removal.
For the decoy control we record both whether the target is spuriously detected and whether the depleted decoy itself is.
Controls~(i) and~(ii) are invariance tests.
They perturb the corpus by the same amount but touch pair~$\cap\,D_q$ only incidentally (an expected $0.06$--$0.19$ documents at $\delta=0.75$), so they test whether $\Delta\text{NPMI}$ reacts to removal in general. Control~(iii) is the specificity test, applying a structurally equivalent \emph{targeted} depletion to another pair.
\end{enumerate}

Table~\ref{tab:removal-driven} reports the outcome.
At $\delta=0.75$, targeted removal newly surfaces 34.1\% (G06N) and 27.3\% (C03C) of pairs, while all three controls stay at or near zero.
Corpus-random and $D_q$-random are at 0\% in every cell but one (a single G06N case, 2.3\%), and depleting a size-matched decoy pair never brings the target into the top-20 (0 of 191 decoy trials across both corpora and all three values of~$\delta$).
The decoy control is the informative one, because it applies the same targeted depletion to a structurally equivalent pair.
In the very same runs the \emph{decoy} enters the top-20 in 25.6--40.5\% (G06N) and 30.0--54.5\% (C03C) of trials.
The detector therefore responds to depletion of whichever pair was depleted, and not to depletion in general. The separation is thus a specificity result, not an artifact of a control that could never succeed.
The recovery rates themselves (34.1\% and 27.3\%) are estimated less precisely than the targeted-versus-control separation. A bootstrap over targets (10{,}000 resamples) gives 95\% intervals of $[20.5\%, 47.7\%]$ (G06N) and $[9.1\%, 45.5\%]$ (C03C), but targets sharing a keyword share a filtered corpus and a top-20 competition, so those intervals understate the variance. Resampling \emph{keyword groups} instead gives $[0.0\%, 41.5\%]$ on G06N ($G=5$ groups, one holding 37 of 44 targets, so a resample can omit it entirely) and $[8.3\%, 42.9\%]$ on C03C ($G=6$). We therefore read the recovery rates as indicative rather than precisely estimated, and rest the claim on the controls being exactly zero.
The effect is non-monotone in $\delta$.
At $\delta=1.0$ the pair's conditional co-occurrence reaches zero and it is excluded by filter~(a) of \S\ref{sec:whitespace} (the total-absence regime), so recovery peaks at partial removal, consistent with BLANC's design target of \emph{underrepresentation} rather than absence.
Because targets, keyword, and ground truth are all fixed before detection, this measure is free of evaluation circularity.

\begin{table*}[t]
\centering
\caption{Removal-driven injection: percentage of established, initially undetected pairs newly entering the top-20 after targeted removal versus three size-matched controls. ``Decoy'' depletes a different established pair in the same $D_q$; its two columns report whether the \emph{target} is spuriously detected (specificity) and whether the depleted \emph{decoy} is (sensitivity). Decoy denominators are smaller where no size-matched decoy exists. Because the decoy arm removes as many documents as the \emph{target} arm, a decoy larger than the target is only partly depleted and can still be recovered, which is why the decoy column stays non-zero at $\delta=1.00$ while the target column is 0\%. Depletion is complete for 22 of 42 G06N decoys and 6 of 20 on C03C at that setting, and none of those fully depleted decoys is recovered, reproducing the total-absence regime on the decoy side.}\label{tab:removal-driven}
\small
\begin{tabular}{lcccccc}
\toprule
 & & & \multicolumn{2}{c}{Random} & \multicolumn{2}{c}{Decoy} \\
\cmidrule(lr){4-5} \cmidrule(lr){6-7}
Corpus & $\delta$ & Targeted & corpus & $D_q$ & target & decoy \\
\midrule
G06N ($n=44$) & 0.50 & 13.6\% & 0\% & 0\% & 0\% & 25.6\% \\
              & 0.75 & \textbf{34.1\%} & 0\% & 0\% & \textbf{0\%} & 40.5\% \\
              & 1.00 & 0\%    & 0\% & 2.3\% & 0\% & 35.7\% \\
\midrule
C03C ($n=22$) & 0.50 & 9.1\%  & 0\% & 0\% & 0\% & 36.4\% \\
              & 0.75 & \textbf{27.3\%} & 0\% & 0\% & \textbf{0\%} & 54.5\% \\
              & 1.00 & 0\%    & 0\% & 0\% & 0\% & 30.0\% \\
\bottomrule
\end{tabular}
\end{table*}

\subsubsection{Baseline comparison}\label{sec:eval-baseline}

We compare BLANC under the identical removal-driven protocol ($\delta=0.75$) against both internal component ablations and external prior methods, replacing one element at a time so each comparison is on equal footing (Table~\ref{tab:baseline}).

\textbf{Multi-view decomposition is essential.}
Replacing the three dimension-specific clusterings with a single BERTopic run on concatenated text (abstract + claims + summary) reproduces the single-view opportunity-discovery setting of Jeon et al.~\cite{Jeon2023}. This yields 0\% recovery on both corpora, because a single view captures each theme within one cluster but cannot express cross-dimensional combinations, leaving no pairs to deplete.

\textbf{NPMI provides specificity; unnormalized connectivity does not.}
Replacing NPMI with Jaccard ($|A\cap B|/|A\cup B|$), cosine similarity, or raw inter-cluster connectivity (the unnormalized co-occurrence count used by the closest precursor, Kim et al.~\cite{Kim2014}), while keeping the multi-view clusters, does \emph{not} cleanly isolate the injected gap.
Although cosine attains a higher \emph{absolute} recovery rate (52.3\% / 45.5\%), every unnormalized or set-normalized measure also ``recovers'' the pair under \emph{random} removal, and the spurious-recovery rate grows as marginal normalization is removed: Jaccard 15.9\% / 27.3\%, cosine 22.7\% / 13.6\%, and inter-cluster connectivity 29.5\% / 54.5\%. For the glass corpus, connectivity recovers the target \emph{more} often under random than under targeted removal.
NPMI is the only measure whose random-removal control is 0\% on both corpora, because its marginal-frequency normalization makes the conditional score react to depletion of the specific combination rather than to any change in cluster sizes.
We therefore characterize NPMI's contribution as \emph{specificity} (zero spurious recovery) rather than raw sensitivity.

\textbf{Learned clustering exposes far more combinations than a standard taxonomy.}
As an external substitute for the learned application axis, we replace it with third-party CPC subgroup codes (\texttt{main\_cpc\_label}), in the spirit of IPC/CPC co-classification gap analysis~\cite{Breschi2003,Curran2011}, keeping the learned novelty clustering and the conditional detector fixed.
The difference is one of \emph{coverage} rather than per-pair recovery.
Substituting CPC codes collapses the number of established cross-dimensional combinations available to test roughly fourfold (high-NPMI pairs fall from 46 to 10 on G06N and from 23 to 11 on C03C), because the off-the-shelf taxonomy aligns poorly with the corpus's actual technology couplings.
On the few combinations CPC does expose, the conditional NPMI detector retains its specificity (targeted recovery 50.0\% and 20.0\% against a 0\% random control), confirming that the specificity property is intrinsic to NPMI and independent of the clustering source, but the candidate pool is too small to support detection on a larger scale.

\begin{table*}[t]
\caption{Baseline comparison on the removal-driven protocol ($\delta=0.75$): percentage of initially undetected target pairs newly entering the top-20 under targeted versus size-matched random removal, the latter drawn from the whole corpus (the \emph{corpus-random} control of \S\ref{sec:eval-removal}). Only NPMI exhibits zero spurious recovery under random removal; a useful detector needs Targeted $\gg$ Random.}\label{tab:baseline}
\centering
\begin{tabular}{lcccc}
\toprule
 & \multicolumn{2}{c}{G06N (ML/AI)} & \multicolumn{2}{c}{C03C (Glass)} \\
\cmidrule(lr){2-3} \cmidrule(lr){4-5}
Method & Targeted & Random & Targeted & Random \\
\midrule
\textbf{Multi-view + NPMI (BLANC)} & \textbf{34.1\%} & \textbf{0.0\%} & \textbf{27.3\%} & \textbf{0.0\%} \\
Multi-view + Jaccard & 29.5\% & 15.9\% & 31.8\% & 27.3\% \\
Multi-view + Cosine  & 52.3\% & 22.7\% & 45.5\% & 13.6\% \\
Multi-view + connectivity~\cite{Kim2014} & 29.5\% & 29.5\% & 22.7\% & 54.5\% \\
Single-view + NPMI~\cite{Jeon2023}   &  0.0\% &  0.0\% &  0.0\% &  0.0\% \\
\bottomrule
\end{tabular}
\end{table*}

All quantitative experiments use publicly available data (HUPD~\cite{Suzgun2024hupd}), enabling full reproducibility; the industrial validation uses proprietary data described in Section~\ref{sec:res-proprietary}.

\section{Discussion}\label{sec:discussion}

\subsection{Practical implications for patent brainstorming}\label{sec:disc-practice}

BLANC turns brainstorming preparation from unstructured reading into a focused, data-driven process.
The three-step workflow (review clusters $\to$ set keyword $\to$ explore white space) runs in minutes once the corpus is prepared, compared with the weeks of manual effort required to build traditional patent maps.
The multi-view decomposition also lets practitioners articulate gaps as specific application--novelty or application--inventive step combinations rather than blank regions on a 2D map.

Keyword selection is a critical user skill.
Domain-specific terms (``antimicrobial'', ``abbe'') substantially outperform generic ones (``glass'', ``layer''), so we recommend starting from Phase-1 cluster keywords, preferring multi-word technical nomenclature, and consulting CPC subclass definitions.
This dependence on expertise does not reintroduce challenge~(2) from \S\ref{sec:intro}.
That challenge was that the \emph{quality of unaided brainstorming} varies with each participant's domain knowledge, leaving the search unstructured and hard to reproduce.
BLANC instead requires expertise only to choose a keyword. Once the keyword is chosen, BLANC structures and accelerates the exploration, makes each query reproducible, and lets practitioners cheaply try several keywords and compare the resulting candidates, turning open-ended reading into a guided, repeatable process rather than removing the need for expertise altogether.

BLANC is a \emph{support tool} rather than a replacement for experts.
Candidates are starting points for discussion, and domain expertise is required to interpret why a given combination is underexplored.
A gap may persist for several reasons: limited technical feasibility, weak market demand, deliberate competitive avoidance, or merely terminological divergence, as in the neuromorphic$\times$spiking gap of \S\ref{sec:res-ws}, where ``neural'' patents simply tend to use different vocabulary. Only the last is a detection artifact, and disentangling a genuine opportunity from these alternatives is precisely where the practitioner's judgment is indispensable.
The proprietary float glass case (\S\ref{sec:res-proprietary}) exemplifies this support role, in that the surfaced candidate matched one that experts had independently identified, focusing their judgment rather than replacing it.

\subsection{Theoretical positioning and generalizability}\label{sec:disc-theory}

The multi-view approach operationalizes recombinant search theory~\cite{Lee2019recombinant} in the patent domain. Its three dimensions of application/use, novelty, and inventive step are complementary facets of technological knowledge, and cross-tabulating them via NPMI reveals recombinant opportunities invisible to single-view analysis.
The NPMI drop metric extends co-word analysis (a bibliometric technique that maps a field through term co-occurrence) from absolute to \emph{conditional} co-occurrence, adding a user-specific lens.
$\Delta\text{NPMI}$ measures the ``surprise'' of a combination's absence in a domain-specific context, paralleling the information-theoretic interpretation of PMI~\cite{Church1990}.

While demonstrated on patents, BLANC is, in principle, applicable to any document collection with multiple classifiable fields.
The cross-domain evaluation supports this.
Despite fundamental technological differences between ML/AI and glass compositions, BLANC shows the same qualitative behavior in the removal-driven test (comparable targeted recovery, 34.1\% vs.\ 27.3\% at $\delta=0.75$, against a 0\% random-removal control in both) with no tuning beyond an HDBSCAN size adjustment, and the same preprocessing patterns proved effective in both domains.

\subsection{Integration with LLMs}\label{sec:disc-llm}

A natural extension is to pass white space candidates to a large language model for explanation generation, articulating \emph{why} a combination is underexplored and \emph{what} approaches might fill it, thus connecting detection to idea generation.
Recent work on patent-specific LLMs~\cite{Lee2020claimgen,Jiang2025claims} suggests this is feasible, and preliminary experiments with commercial LLMs yield plausible explanations from cluster-keyword and co-occurrence context, though systematic evaluation remains future work.

\subsection{Limitations}\label{sec:disc-limit}

Several limitations deserve mention.
\emph{Statistical stability}: when the filtered corpus is small, NPMI estimates become unstable, and minimum-occurrence thresholds only partially mitigate this; we report bootstrap confidence intervals on the removal-driven recovery rate (\S\ref{sec:eval-removal}), but these intervals are wide given the modest number of eligible targets, and the experiments confirm that pairs with very small co-occurrence counts ($\leq 12$) are more likely to be missed.
\emph{Preprocessing and dimension choice}: multi-view clustering depends on structured text fields, and patent claims in particular require domain-aware boilerplate removal (\S\ref{sec:eval-sensitivity}); the three-dimension decomposition (application, novelty, inventive step) is also not the only possible choice, and a formal sensitivity analysis to the number of views $K$ and to alternative dimensions (e.g., technical field, problem addressed) remains future work.
\emph{Objectivity of cluster estimation}: we prefer automatic HDBSCAN cluster estimation precisely because it avoids analyst-chosen cluster counts; on the small proprietary corpus, however, the novelty dimension fell back to a manually specified $k$-means ($k=10$) because HDBSCAN underclustered that field, which lowers the objectivity of that one case and argues for automatic estimation as the default.
\emph{Keyword and temporal dynamics}: the system does not automatically suggest optimal keywords (Appendix~\ref{sec:eval-synthetic}) and does not account for temporal dynamics; periodic re-analysis is recommended.
\emph{Nature of the evaluation}: because white space is defined by absence, no objective ground truth exists, so synthetic injection and the single expert-confirmed industrial case serve only as proxies for true discovery performance.
Synthetic injection in particular detects the \emph{reduction} of known cluster pairs rather than the emergence of genuinely novel combinations; truly absent combinations can instead be identified by direct inspection of the co-occurrence tensor.
\emph{Independence across targets}: the keyword rule of \S\ref{sec:eval-removal} is deterministic but not injective, and 37 of the 44 G06N targets share the keyword ``computer''. The targets are therefore conditioned on a handful of terms rather than being independent queries, which is why we report keyword-group as well as target-level bootstrap intervals there and treat the recovery rates as indicative.
\emph{Regime covered by the evaluation}: the removal-driven protocol selects deliberately broad keywords, so its filtered corpora are large (a median 76\% of the corpus on G06N and 55\% on C03C). Intended use, by contrast, is narrow: ``neural'' retains 18.7\% of G06N and ``fluorine'' 4.6\% of the proprietary corpus. The primary evaluation thus does not probe the small-$D_q$ regime where NPMI estimates are least stable (\S\ref{sec:disc-limit}, statistical stability). Running the same protocol with narrow keywords does not solve this problem, because the pair would often be detected before removal, reinstating the circularity the protocol was designed to avoid. That regime is covered only by the diagnostic of Appendix~\ref{sec:eval-synthetic} and the industrial case of \S\ref{sec:res-proprietary}.
\emph{Scope of industrial validation}: the proprietary-data validation (\S\ref{sec:res-proprietary}) rests on a single expert-confirmed candidate; broader validation across additional fields and expert panels would strengthen the evidence.

\section{Conclusion}\label{sec:conclusion}

We have presented BLANC, a three-phase pipeline for patent white space detection that integrates multi-view BERTopic clustering, NPMI-based co-occurrence analysis, and conditional gap detection via a new NPMI drop ($\Delta\text{NPMI}$) metric.
BLANC identifies technology combinations that are well-established in the general patent landscape but underexplored for a user-specified keyword, presenting them as actionable candidates for new patent applications.

A removal-driven injection evaluation on two public USPTO corpora, comprising 5{,}417 ML/AI patents in CPC G06N and 1{,}982 glass composition patents in CPC C03C, tests whether BLANC recovers artificially depleted technology combinations under a protocol conditioned on a keyword independent of the target.
When three-quarters of a target pair's documents are removed, BLANC recovers 34.1\% (G06N) and 27.3\% (C03C) of them, while size-matched control removals essentially never do.
Depleting a different established combination never brings the target into the top-20 (0 of 191 trials), while random removals leave it undetected in 11 of 12 cells.
Comparisons on the same protocol against component ablations and external prior methods establish three things: multi-view decomposition is indispensable (single-view 0\%); among all association measures tested, including the closest precursor (unnormalized inter-cluster connectivity), NPMI is the only one with zero spurious recovery under random removal; and the learned application clustering exposes several times as many established cross-dimensional combinations as a standard CPC taxonomy.
The consistency of this behavior across two technologically distinct domains supports the generalizability of the framework.
A proprietary-data case on 302~float glass / glass-ceramics patents further illustrates that, with the keyword ``fluorine,'' BLANC reveals a fluorine surface treatment $\times$ warpage suppression candidate ($\Delta\text{NPMI}$ up to 0.48) that IP experts had independently identified by hand---a single confirmatory case.

A key practical finding is that field-specific text preprocessing is essential for effective multi-view clustering; without it, dimensions collapse into uninformative clusters.
Future work includes temporal analysis of white space evolution, multi-keyword conditioning, statistical confidence scoring via bootstrap, and LLM-based explanation generation for white space candidates.

\section*{Data availability}
Quantitative experiments use the publicly available Harvard USPTO Patent Dataset (HUPD)~\cite{Suzgun2024hupd}, accessible at \url{https://huggingface.co/datasets/HUPD/hupd}.
The proprietary float glass patent data used for industrial validation were retrieved from a commercial patent database under institutional license; neither the records nor the underlying search query can be redistributed. The corpus composition (size, field structure, and filing window) is described in Section~\ref{sec:res-proprietary}.

\section*{Funding}
This research did not receive any specific grant from funding agencies in the public, commercial, or not-for-profit sectors.

\section*{Declaration of competing interest}
The authors declare that they have no known competing financial interests or personal relationships that could have appeared to influence the work reported in this paper.
Part of the methodology described in this paper is the subject of a Japanese patent application (Publication No.\ JP2025148721A; Miyazawa, S., Fujii, K., Fujii, Y., Tomiyori, Y., ``Information processing device, information processing method, and program,'' filed 2024-03-26, published 2025-10-08; Assignee: AGC Inc.).

\section*{CRediT authorship contribution statement}
\textbf{Shuichi Miyazawa:} Conceptualization, Methodology, Software, Validation, Visualization, Writing -- original draft.
\textbf{Kensuke Fujii:} Conceptualization, Validation, Writing -- review \& editing.

\section*{Declaration of generative AI and AI-assisted technologies in the manuscript preparation process}
During the preparation of this work the authors used Claude Code (Anthropic) to assist with language editing, restructuring of paragraphs, and consistency checks across sections of the manuscript.
After using this tool, the authors reviewed and edited the content as needed and take full responsibility for the content of the publication.

\appendix
\section{Public data experiment parameters}\label{app:public}

Both corpora use the Harvard USPTO Patent Dataset (HUPD)~\cite{Suzgun2024hupd}.

All experiments were implemented in Python using BERTopic, Sentence-Transformers, UMAP, and HDBSCAN.
Because the clustering relies on these libraries' numerical routines, the exact NPMI values are sensitive to their versions, whereas the rank-based recovery results of \S\ref{sec:eval-removal} are robust.
All stochastic components (UMAP, $k$-means, and the injection-experiment random number generator) use fixed random seeds (HDBSCAN is deterministic given its inputs), so the reported numbers are deterministic for a given corpus and configuration.

\textbf{Numerical stability.} The small constant added to each probability is $\varepsilon = 10^{-8}$ for the cluster-pair NPMI of Eqs.~(\ref{eq:marginal})--(\ref{eq:npmi}), and $\varepsilon = 10^{-12}$ for the term--cluster NPMI in the keyword score of Eq.~(\ref{eq:score}).

\subsection*{G06N (ML/AI) corpus}
5{,}417 patent applications with CPC subclass G06N, filed 2013--2016.

\textbf{Text field mapping:}
\begin{itemize}
\item Application/use dimension: \texttt{abstract}
\item Novelty dimension: \texttt{claims} (with legal boilerplate removed)
\item Inventive step dimension: \texttt{summary} (with figure descriptions and boilerplate removed)
\end{itemize}

\textbf{Claims preprocessing} removes, via regular expressions, claim numbering (``1.\ A method\ldots''), back-references (``The method of claim~1''), and common formulaic patterns (``comprising'', ``wherein'', ``configured to'', ``non-transitory computer-readable medium'', etc.).
Effect: novelty clustering improved from 2~clusters to 47~clusters.

\textbf{Summary preprocessing} removes section headers (the \texttt{<SOH>\allowbreak\ldots<EOH>} markup that delimits headings in HUPD's summary field), entire FIG.\ sentences, and common formulaic patterns (``in one embodiment'', ``the present disclosure'', etc.).
Effect: inventive step clustering improved from 8~clusters (80\% in one dominant cluster) to 51~clusters.

\textbf{Cluster keyword scoring.}
Rather than BERTopic's built-in c-TF-IDF, which favors high-frequency terms in specialized patent text, each cluster's representative keywords are ranked by a composite score that balances within-cluster frequency and corpus-level distinctiveness, weighted by a cluster-level inverse document frequency (cluster-IDF):
\begin{equation}\label{eq:score}
\mathrm{score}(w, i) = \bigl[\alpha \cdot \mathrm{NPMI}(w, i) + (1 - \alpha) \cdot P(w \mid i)\bigr] \cdot \bigl(1 + \mathrm{cIDF}(w)\bigr)
\end{equation}
where $P(w \mid i)$ is the fraction of documents in cluster $i$ containing term $w$, $\mathrm{NPMI}(w, i)$ is the normalized PMI between term and cluster, $\alpha \in [0,1]$ is a mixing weight (default 0.5), and $\mathrm{cIDF}(w) = \log (N_{\mathrm{c}} / |\{i : w \in \mathrm{vocab}(i)\}|)$ is the cluster-IDF with $N_{\mathrm{c}}$ the total number of clusters in the dimension (distinct from the number of views $K$).
A term appearing in every cluster receives $\mathrm{cIDF}=0$; a term unique to one cluster receives $\mathrm{cIDF}=\log N_{\mathrm{c}}$, substantially boosting its rank, which suppresses legal boilerplate and structural terms such as ``comprising'' and ``plurality'' that otherwise dominate keyword lists.

\textbf{Greedy diversified keyword selection.}
Candidates ranked by Eq.~(\ref{eq:score}) are chosen sequentially with three rules:
(1)~if a candidate's stemmed tokens are a subset of an already-selected longer term, skip;
(2)~if a candidate is a longer phrase that subsumes an already-selected shorter term, it \emph{replaces} the shorter term (e.g., ``acid compound'' replaces ``acid'');
(3)~otherwise skip if more than half of the candidate's stemmed tokens already appear among selected keywords.
All overlap checks use Snowball (Porter2)~\cite{Porter2001}, a word-stemming algorithm, so that inflectional variants (``generate''/``generating'') are recognized as redundant.

\textbf{BERTopic parameters:}
\begin{itemize}\setlength{\itemsep}{0pt}
\item Embedding: \texttt{all-MiniLM-L6-v2} (384-dim)
\item UMAP: \texttt{n\_neighbors=15}, \texttt{n\_components=5}, \texttt{min\_dist=0}
\item HDBSCAN: \texttt{min\_cluster\_size=20}, \texttt{min\_samples=5}
\item Keywords: $\alpha=0.5$, cluster-IDF, Snowball, top-10
\item White space detection: $\theta_{\min}=0.3$, top-20
\item Stop words: standard English list augmented with patent-domain terms
\end{itemize}

\textbf{Clustering results (after preprocessing):}
Application dimension: 62 clusters (noise rate 36.8\%);
Novelty dimension: 47 clusters (noise rate 33.2\%);
Inventive step dimension: 51 clusters (noise rate 28.3\%).
The application$\times$novelty combination yields $62 \times 47 = 2{,}914$ cluster pairs, of which 58 have NPMI~$\geq 0.3$ and co-occurrence~$\geq 8$ (46 at the stricter $\geq 12$ co-occurrence floor used by the removal-driven evaluation, \S\ref{sec:eval-removal}).

\subsection*{C03C glass corpus}

1{,}982 patent applications with CPC subclass C03C (chemical composition of glasses, glazes, or vitreous enamels), filed 2012--2017.
The same text field mapping and claims preprocessing are applied; the summary (inventive step) field is left unprocessed, as the glass-domain vocabulary clusters well without boilerplate removal.
BERTopic parameters: identical to G06N except HDBSCAN \texttt{min\_cluster\_size=10}, \texttt{min\_samples=3} (adjusted for the smaller corpus).
Clustering results: 63, 63, and 48 clusters (noise rates 21.3\%, 21.3\%, 17.3\%).
The application$\times$novelty combination yields 59 pairs with NPMI~$\geq 0.3$ and co-occurrence~$\geq 5$ (23 at the $\geq 12$ floor used by the removal-driven evaluation).
The application$\times$inventive step combination yields 54 pairs with NPMI~$\geq 0.3$ and co-occurrence~$\geq 5$.

\section{Theme-keyword sensitivity diagnostic}\label{sec:eval-synthetic}

Our initial protocol paired each target with a keyword \emph{semantically related} to it: given a cluster pair $(c_a, c_b)$ with high global NPMI, remove a fraction $\delta$ of its documents, recompute NPMI on the reduced corpus, and run white space detection with a related keyword $q$; a successful detection places the target in the top-20.
We repeat this for 16 diverse themes per corpus on both HUPD corpora under identical protocols, with \emph{Hit Rate} the fraction of experiments with the target in the top-20, \emph{MRR} the mean of $1/\mathrm{rank}$, and \emph{Precision@$k$} the fraction within rank~$k$.
We retain this protocol as a \emph{diagnostic} of $\Delta$NPMI's sensitivity, not as a performance claim; the controls at the end of this section show why a pair-correlated keyword cannot, on its own, evidence gap recovery, which is why the removal-driven evaluation (\S\ref{sec:eval-removal}) is our primary quantitative result.

\textbf{Theme selection procedure.}
To avoid arbitrary selection, we rank all cluster pairs in the application$\times$novelty dimension by global NPMI (with a minimum co-occurrence filter of 8 for G06N / 5 for C03C) and go through the ranked list from the top, selecting pairs that admit an unambiguous domain-specific keyword and are not semantically redundant with an already-selected pair; selection stops at 16 pairs.
On G06N this covers ranks 1--20 (4 skipped); on C03C, ranks 1--25 (9 skipped).
Skipped pairs typically involve generic descriptors (e.g., ``cognitive'', ``coating'') that cannot be targeted by a single keyword, or that duplicate the semantic content of a higher-ranked pair.

\textbf{Results.}
Tables~\ref{tab:synthetic} and~\ref{tab:synthetic-glass} give per-theme outcomes.
On G06N ($\delta=0.50$), BLANC achieves Hit Rate 93.8\% with 5 of 16 targets at rank~1, MRR 0.469, P@5 68.8\%; the one miss (ontology) has a keyword that matches documents across many cluster pairs, diluting the signal.
On C03C with domain-specific keywords (``antimicrobial'', ``abbe'', ``optical fiber''), BLANC achieves Hit Rate 81.2\% and MRR 0.594; the three misses involve cluster pairs with very small co-occurrence counts ($\leq 12$).

\begin{table*}[!htbp]
\centering
\begin{minipage}[t]{0.48\textwidth}
\centering
\captionof{table}{Synthetic injection results on G06N ($\delta=0.50$, app$\times$novelty).}\label{tab:synthetic}
\small
\begin{tabular}{llrcr}
\toprule
Theme & KW & Rem & Rank & $\Delta$ \\
\midrule
Quantum comp.      & quantum    & 54 & 1  & 0.733 \\
Autonomous drv.     & vehicle    & 21 & 1  & 0.463 \\
Anomaly det.        & anomaly    & 19 & 1  & 0.404 \\
QA system           & question   & 48 & 1  & 0.247 \\
Finite automaton    & automaton  & 16 & 1  & 0.103 \\
Sentiment           & sentiment  & 10 & 2  & 0.359 \\
Rule engine         & rule       & 28 & 3  & 0.188 \\
Recommendation      & recommend  & 26 & 3  & 0.135 \\
Multi-obj opt.      & optimization & 29 & 3 & 0.097 \\
Time series         & forecast   & 25 & 4  & 0.061 \\
Energy cons.        & energy     & 32 & 4  & 0.023 \\
Graph NN            & graph      & 15 & 6  & 0.187 \\
Social network      & social     & 33 & 8  & 0.129 \\
Decision tree       & dec.\ tree & 20 & 8  & 0.051 \\
Predictive model    & predictive & 11 & 11 & 0.124 \\
Ontology            & ontology   & 20 & MISS & --- \\
\midrule
\multicolumn{3}{l}{Hit Rate: 15/16 (93.8\%)}  & \multicolumn{2}{l}{MRR: 0.469} \\
\multicolumn{5}{l}{P@1: 31.2\% / P@5: 68.8\% / P@10: 87.5\%} \\
\bottomrule
\end{tabular}
\end{minipage}
\hfill
\begin{minipage}[t]{0.48\textwidth}
\centering
\captionof{table}{C03C injection ($\delta=0.50$, app$\times$nov, domain KWs).}\label{tab:synthetic-glass}
\small
\begin{tabular}{llrcr}
\toprule
Theme & KW & Rem & Rank & $\Delta$ \\
\midrule
Antimicrobial       & antimicrobial & 9 & 1 & 0.592 \\
Solar glazing       & glazing       & 12 & 1 & 0.252 \\
FTIR optical        & ftir          &  7 & 1 & 0.803 \\
Solar transm.       & dom.\ wavel. & 15 & 1 & 0.620 \\
Optical fiber       & opt.\ fiber  & 14 & 1 & 0.800 \\
Sizing comp.        & sizing        &  6 & 1 & 0.515 \\
Li silicate         & Li silicate   & 19 & 2 & 0.102 \\
Silica-titania      & titania       &  4 & 2 & 0.174 \\
Glass fiber         & fiber comp.   & 20 & 2 & 0.212 \\
Optical (Abbe)      & abbe          & 22 & 2 & 0.709 \\
Tempered glass      & tempered      &  9 & 2 & 0.162 \\
Etching             & etching       & 13 & 2 & 0.262 \\
Induction seal      & sealing       &  7 & 2 & 0.430 \\
LCD matrix          & liq.\ crystal &  6 & MISS & --- \\
Windshield          & windshield    &  5 & MISS & --- \\
Deep compr.         & compr.\ str. &  9 & MISS & --- \\
\midrule
\multicolumn{3}{l}{Hit Rate: 13/16 (81.2\%)}  & \multicolumn{2}{l}{MRR: 0.594} \\
\multicolumn{5}{l}{P@1: 37.5\% / P@3: 81.2\% / P@5: 81.2\%} \\
\bottomrule
\end{tabular}
\end{minipage}
\end{table*}

\textbf{Third dimension validation.}
Repeating the injection on the application$\times$inventive step pair confirms the diagnostic transfers to the third dimension: Hit Rate 87.5\% (MRR 0.457) on C03C. The same holds on G06N once the summary field receives field-specific preprocessing, which raises the inventive-step dimension from 8 collapsed clusters to 51 and restores detection.

\textbf{Impact of keyword specificity.}
Rerunning the 16 glass themes with an \emph{automatically generated} keyword (the first word of the corresponding application cluster's top keyword) drops Hit Rate to 56.2\% (MRR 0.287) from 81.2\% (0.594) with domain-specific keywords, because generic terms (``glass'', ``layer'') match many cluster pairs and dilute the conditional signal.
BLANC therefore performs best with specific technical terms; the generalizability claim itself rests on the removal-driven evaluation (\S\ref{sec:eval-removal}), which shows the same targeted-versus-random behavior in both corpora.

\textbf{Removal fraction sweep.}
We sweep $\delta \in \{0.10,\allowbreak 0.25,\allowbreak 0.50,\allowbreak 0.75,\allowbreak 0.90,\allowbreak 1.00\}$ on the quantum-computing pair (application cluster~2 $\times$ novelty cluster~4, global NPMI of 0.978, 106 documents) in G06N (Table~\ref{tab:delta-sweep}).
BLANC detects the target at rank~1 for every $\delta < 1.0$, with $\Delta\text{NPMI}$ rising monotonically from 0.726 to 0.786.
At $\delta=1.0$ (complete removal), the target is excluded by the filtering criterion (\S\ref{sec:whitespace}) because conditional co-occurrence drops to zero. BLANC is designed for \emph{underrepresentation}, not \emph{total absence}.

\begin{table}[!htbp]
\centering
\caption{$\delta$ sweep for the quantum-computing pair (app$\times$nov). Removal counts are taken over all 109~documents assigned to the pair in the application and novelty dimensions. The co-occurrence count of 106 in Table~\ref{tab:npmi-top} is smaller because three of these documents are unassigned (noise) in the inventive step dimension and are thus excluded from the tensor (\S\ref{sec:tensor}).}\label{tab:delta-sweep}
\begin{tabular}{cccc}
\toprule
$\delta$ & Removed & Rank & $\Delta$NPMI \\
\midrule
0.10 &  10 & 1 & 0.726 \\
0.25 &  27 & 1 & 0.728 \\
0.50 &  54 & 1 & 0.733 \\
0.75 &  81 & 1 & 0.747 \\
0.90 &  98 & 1 & 0.786 \\
1.00 & 109 & MISS & --- \\
\bottomrule
\end{tabular}
\end{table}

\textbf{Controls: what document removal contributes.}
The protocol above pairs each target with a keyword semantically related to it, so we add two controls to test whether detection is driven by the injection or by the keyword.
Under a \emph{no-removal} baseline (the keyword applied to the unperturbed corpus), the hand-selected target already appears in the top-20 for 100\% (16/16, G06N) and 87.5\% (14/16, C03C) of themes, and a \emph{size-matched random-removal} control (removing the same number of randomly chosen documents) reproduces these rates exactly.
Removing the target pair's documents lowers the hit rate only slightly (G06N 93.8\%, C03C 81.2\%, the latter agreeing with Table~\ref{tab:synthetic-glass}) and shifts only a few targets between adjacent ranks.
The reason is structural.
Conditioning on a keyword that correlates with a pair's clusters inflates those clusters' marginal probabilities within the small filtered corpus $D_q$, which deflates the conditional NPMI through the normalization in Eq.~(\ref{eq:npmi}) even when the conditional co-occurrence count is undiminished.
The hit rates of Tables~\ref{tab:synthetic}--\ref{tab:synthetic-glass} therefore measure BLANC's intended \emph{production} behavior (surfacing the cluster pair associated with a keyword) rather than recovery of an artificially injected gap.

\section{Keyword assignment in the removal-driven protocol}\label{app:keywords}

The rule of Step~2 in \S\ref{sec:eval-removal} is deterministic but not injective, so a few generic terms cover all targets. Table~\ref{tab:keywords} lists every assigned keyword with the size of the corpus it selects and the share of that corpus occupied by the target pairs it serves. The shares are uniformly small, at most 7.0\%, which is what makes the pairs undetectable before removal; the concentration on ``computer'' is the basis of the independence caveat in \S\ref{sec:disc-limit}.

\begin{table}[h]
\centering
\caption{Keywords assigned by Step~2 of \S\ref{sec:eval-removal}. Pairs: targets receiving the keyword. $|D_q|$: documents selected by it. Share: $|\text{pair}\cap D_q|/|D_q|$ over those targets.}\label{tab:keywords}
\small
\begin{tabular}{llrrrr}
\toprule
 & & & & \multicolumn{2}{c}{Share of $D_q$} \\
\cmidrule(lr){5-6}
Corpus & Keyword & Pairs & $|D_q|$ & median & max \\
\midrule
G06N & computer    & 37 & 4{,}118 & 0.56\% & 2.21\% \\
     & data        &  4 & 3{,}789 & 0.50\% & 0.69\% \\
     & information &  1 & 2{,}417 & 0.54\% & 0.54\% \\
     & memory      &  1 & 2{,}307 & 1.34\% & 1.34\% \\
     & state       &  1 &   984 & 7.01\% & 7.01\% \\
\midrule
C03C & surface     &  8 & 1{,}159 & 1.51\% & 1.64\% \\
     & temperature &  7 & 1{,}086 & 1.29\% & 3.31\% \\
     & layer       &  3 &   966 & 2.07\% & 2.59\% \\
     & sio2        &  2 &   879 & 2.45\% & 3.30\% \\
     & composition &  1 &   863 & 1.27\% & 1.27\% \\
     & forming     &  1 &   756 & 1.72\% & 1.72\% \\
\bottomrule
\end{tabular}
\end{table}

\bibliographystyle{model1-num-names}
\bibliography{references}

\end{document}